\documentclass[aps,twocolumn,showpacs,byrevtex,prl,reprint]{revtex4-1}
\RequirePackage{lineno}

\usepackage{graphicx}
\usepackage{dcolumn}
\usepackage{bm}
\usepackage{rotating}
\usepackage{epstopdf}
\usepackage{color}
\usepackage{verbatim} 
\usepackage{multirow}
\usepackage[abs]{overpic}
\usepackage{amsmath}
\usepackage{amssymb}
\usepackage{xspace}
\usepackage[titletoc]{appendix}
\usepackage{float}

\usepackage[colorlinks,
            linkcolor=blue,
            anchorcolor=blue,
            citecolor=blue]{hyperref}

\newcommand{\PreserveBackslash}[1]{\let\temp=\\#1\let\\=\temp}
\newcolumntype{C}[1]{>{\PreserveBackslash\centering}p{#1}}
\newcolumntype{R}[1]{>{\PreserveBackslash\raggedleft}p{#1}}
\newcolumntype{L}[1]{>{\PreserveBackslash\raggedright}p{#1}}

\newcommand{\EE}{e^+e^-}

\newcommand{\too}{\rightarrow}

\begin{document}
\graphicspath{{figure/}}
\DeclareGraphicsExtensions{.eps,.png,.ps}
\title{\boldmath First measurement of $\Sigma^{+}n\too\Lambda p$ and $\Sigma^{+}n\too\Sigma^{0}p$ cross sections via $\Sigma^+$-nucleus scattering at an electron-positron collider}
\author{
  \begin{small}
    \begin{center}
      M.~Ablikim$^{1}$, M.~N.~Achasov$^{4,c}$, P.~Adlarson$^{76}$, X.~C.~Ai$^{81}$, R.~Aliberti$^{35}$, A.~Amoroso$^{75A,75C}$, Q.~An$^{72,58,a}$, Y.~Bai$^{57}$, O.~Bakina$^{36}$, Y.~Ban$^{46,h}$
% ...
\\
\vspace{0.2cm}
(BESIII Collaboration)\\
\vspace{0.2cm} {\it
$^{1}$ Institute of High Energy Physics, Beijing 100049, People's Republic of China\\
% ...
$^{4}$ Budker Institute of Nuclear Physics SB RAS (BINP), Novosibirsk 630090, Russia\\
% ...
$^{35}$ Johannes Gutenberg University of Mainz, Johann-Joachim-Becher-Weg 45, D-55099 Mainz, Germany\\
$^{36}$ Joint Institute for Nuclear Research, 141980 Dubna, Moscow region, Russia\\
% ...
$^{46}$ Peking University, Beijing 100871, People's Republic of China\\
% ...
$^{57}$ Southeast University, Nanjing 211100, People's Republic of China\\
$^{58}$ State Key Laboratory of Particle Detection and Electronics, Beijing 100049, Hefei 230026, People's Republic of China\\
% ...
$^{72}$ University of Science and Technology of China, Hefei 230026, People's Republic of China\\
% ...
$^{75}$ University of Turin and INFN, (A)University of Turin, I-10125, Turin, Italy; (B)University of Eastern Piedmont, I-15121, Alessandria, Italy; (C)INFN, I-10125, Turin, Italy\\
$^{76}$ Uppsala University, Box 516, SE-75120 Uppsala, Sweden\\
% ...
$^{81}$ Zhengzhou University, Zhengzhou 450001, People's Republic of China\\
\vspace{0.2cm}
$^{a}$ Deceased\\
% ...
$^{c}$ Also at the Novosibirsk State University, Novosibirsk, 630090, Russia\\
% ...
$^{h}$ Also at State Key Laboratory of Nuclear Physics and Technology, Peking University, Beijing 100871, People's Republic of China\\
% ...
      }\end{center}
    \vspace{0.4cm}
\end{small}
}
\affiliation{}


\begin{abstract}
Using $(1.0087\pm0.0044)\times10^{10}$ $J/\psi$ events collected with the BESIII detector at the BEPCII storage ring, the reactions $\Sigma^{+}n\rightarrow\Lambda p$ and $\Sigma^{+}n\rightarrow\Sigma^{0}p$ are studied, where the $\Sigma^{+}$ baryon is produced in the process $J/\psi\rightarrow\Sigma^{+}\bar{\Sigma}^-$ and the neutron is a component of the $^9\rm{Be}$, $^{12}\rm{C}$ and $^{197}\rm{Au}$ nuclei in the beam pipe. Clear signals of these two reactions are observed for the first time. Their cross sections are measured to be $\sigma(\Sigma^{+}+{^9\rm{Be}}\rightarrow\Lambda+p+X)=(45.2\pm12.1_{\rm{stat}}\pm7.2_{\rm{sys}})$~mb and $\sigma(\Sigma^{+}+{^9\rm{Be}}\rightarrow\Sigma^{0}+p+X)=(29.8\pm9.7_{\rm{stat}}\pm6.9_{\rm{sys}})$~mb for a $\Sigma^{+}$ average momentum of $0.992$~GeV/$c$, within a range of $\pm0.015$~GeV/$c$, where $X$ represents the residual nucleus. This is the first study of $\Sigma^{+}$-nucleon scattering at an electron-positron collider.
\end{abstract}

\maketitle

Measurements of hyperon-nucleon (YN) scattering have been performed for more than half a century, starting in the  1960s~\cite{introduction1, introduction2, introduction3, introduction4, introduction5, introduction6, introduction7, introduction8, introduction9}. However, due to the limited availability of hyperon beams, which are  also characterized by short lifetimes and low intensities, experimental data on YN interactions are scarce. This lack of data restricts theoretical progress on YN interactions, which is essential for understanding the nature of neutron stars (NS)~\cite{neutronstar1, neutronstar2, neutronstar3, neutronstar4}. In the inner core of NS, it is believed that hyperons may exist because the nucleon chemical potential is sufficiently high to make the conversion of neutrons into hyperons energetically favorable. However, this conversion decreases the Fermi pressure of the system in the equation of state (EoS), thus reducing the mass that NS can sustain.  EoS calculations that include hyperons often predict a maximum mass of NS lower than that which is observed, a discrepancy known as the ``hyperon puzzle"~\cite{neutronstar1, neutronstar2, neutronstar3, neutronstar4}. It has been suggested that this problem could be overcome through the inclusion of a missing three-body hyperon-nucleon-nucleon (YNN) interaction, which arises naturally from $\Lambda$N-$\Sigma$N coupling~\cite{lambda-sigma1, lambda-sigma2}. Therefore, acquiring more YN scattering data is crucial for refining these calculations and providing constraints on the hyperon puzzle in the context of neutron stars.

The $\Lambda$N-$\Sigma$N coupling effect is also believed to be significant in  $\Lambda$ and $\Sigma$ hypernuclei~\cite{lambda-sigma3, lambda-sigma4, lambda-sigma5, lambda-sigma6}, highlighting the critical importance of understanding the strength of this coupling for elucidating the nature of these systems. Furthermore, it has been a long-standing puzzle that the reported charge symmetry breaking (CSB) effects in the $\Lambda$N interaction are relatively large~\cite{csb1, csb2, csb3}. An important contribution to this CSB is attributed to $\Lambda$-$\Sigma$ mixing, resulting in a one-pion exchange contribution to the $\Lambda$N interaction. Precise measurements of $\Lambda$-$\Sigma$ conversion, such as the reaction $\Sigma^{+}n\rightarrow\Lambda p$, are essential for addressing these issues. The $\Sigma$-nucleon scattering has been studied in different theoretical models~\cite{theory1, theory2, theory3, theory4, theory5, theory6, theory7, theory8, theory9, theory10}, such as the constituent quark model~\cite{theory1}, the meson-exchange picture~\cite{theory2, theory3, theory4, theory5}, and the chiral effective field theory approach~\cite{theory6, theory7, theory8}. Despite several measurement campaigns of $\Sigma$-nucleon scattering~\cite{sigma1, sigma2, sigma3, sigma4, sigma5, sigma6, sigma7, sigma8}, experimental knowledge is still limited. Therefore, there is an urgent need for new experimental measurements on $\Sigma$-nucleon scattering to validate and refine these theoretical models, thereby advancing research in this field.

The BESIII detector records symmetric $e^+e^-$ collisions at the BEPCII collider~\cite{bepcii}. Details of the BESIII detector can be found in Ref.~\cite{besiii}. The material of the beam pipe is composed of gold ($^{197}\rm{Au}$), beryllium ($^{9}\rm{Be}$) and oil $(^{12}\rm{C}:$$^{1}\rm{H}$$=1:2.13)$~\cite{xiscatteringbes, lambdabarscatteringbes, lambdascatteringbes}. With a sample of $(1.0087\pm0.0044)\times10^{10}$ $J/\psi$ events~\cite{totalnumber} collected by the BESIII detector, almost monoenergetic $\Sigma^{+}$ hyperons, which can serve as an intense hyperon beam, are produced via the decay $J/\psi\too\Sigma^{+}\bar{\Sigma}^{-}$, with a momentum of $(0.992\pm0.015)$~GeV/$c$, where the momentum spread results from the small horizontal crossing angle of $11$~mrad for the $e^{\pm}$ beams. The $\Sigma^{+}$ baryons can interact with the material in the beam pipe, allowing for the measurement of $\Sigma^{+}$-nucleon scattering. Studies of the similar $\Xi^0/\Lambda/\bar{\Lambda}$-nucleon scattering processes  using this novel method have been reported by  BESIII~\cite{xiscatteringbes, lambdabarscatteringbes, lambdascatteringbes}.

In this Letter, we present a study of the reactions $\Sigma^{+}n\too\Lambda p$ and $\Sigma^{+}n\too\Sigma^{0}p$. The $\Sigma^{+}$ is produced in the process $J/\psi\too\Sigma^{+}\bar{\Sigma}^{-}$, while the neutron originates from a component of the $^9\rm{Be}$, $^{12}\rm{C}$ and $^{197}\rm{Au}$ nuclei in the beam pipe. This study constitutes the first investigation of $\Sigma$-nucleon interactions at an electron-positron collider. The cross-sections of the reactions $\Sigma^{+}+{^9\rm{Be}}\too\Lambda+p+X$ and $\Sigma^{+}+{^9\rm{Be}}\too\Sigma^{0}+p+X$ are also determined. Since the momentum of the incident $\Sigma^+$ is relatively high, its interaction with nuclei tends to be a direct nuclear reaction. It is assumed that a $\Sigma^+$ reacts with a single neutron in the $^9\rm{Be}$ directly in the reaction, giving rise to a residual nucleus $X$, a $\Lambda/\Sigma^{0}$ and a proton in the final state. In determining the cross-sections of $\Sigma^{+}+{^9\rm{Be}}\too\Lambda+p+X$ and $\Sigma^{+}+{^9\rm{Be}}\too\Sigma^{0}+p+X$ from the composite material, the reactions are assumed to be pure surface processes~\cite{ratio1}.

In this analysis, simulated data samples that are generated with a {\sc{Geant4}}-based Monte Carlo (MC) package~\cite{geant4}, which includes the geometric description of the BESIII detector~\cite{display} and the detector response, are used to determine detection efficiencies and to estimate backgrounds. The inclusive MC sample includes both the production of the $J/\psi$ resonance and the continuum processes incorporated in {\sc kkmc} package~\cite{KKMC}. All particle decays are modeled with {\sc evtgen}~\cite{ref:evtgen} using branching fractions either taken from the Particle Data Group (PDG)~\cite{pdg}, where available, or otherwise estimated with {\sc lundcharm}~\cite{ref:lundcharm}. Final-state radiation from charged final-state particles is incorporated using the {\sc photos} package~\cite{photos}.

The signal processes considered in this analysis are $J/\psi\too\Sigma^+\bar{\Sigma}^-$, $\Sigma^{+}n\too\Lambda p$ and $\Sigma^{+}n\too\Sigma^{0}p$, with $\Sigma^{0}\too\gamma\Lambda$, $\Lambda\too p\pi^-$, $\bar{\Sigma}^{-}\too\bar{p}\pi^0$, $\pi^0\too\gamma\gamma$. In order to determine the detection efficiency, the signal MC events are simulated with the angular distribution of $J/\psi\too\Sigma^+\bar{\Sigma}^-$ generated according to the measurement in Ref.~\cite{alpha}. We simulate the reactions $\Sigma^{+}n\too\Lambda p$ and $\Sigma^{+}n\too\Sigma^{0}p$ assuming the neutron to be free, regardless of its Fermi-momentum. Since the momentum of the incident $\Sigma^{+}$ is much greater than the Fermi-momentum, which is about one hundred~MeV/$c$, this approximation is reasonable, and the effect of this approximation is considered in the systematic-uncertainty evaluation. We tune the center-of-mass energy of the reaction system in the MC simulation to make the $M(\Lambda p)$ and $M(\Sigma^{0} p)$ distributions consistent with those in data. The angular distributions of the reactions are generated using an isotropic phase-space distribution.

Charged tracks detected in the multilayer drift chamber (MDC) are required to lie within a polar-angle ($\theta$) range of $|\!\cos\theta|<0.93$, where $\theta$ is defined with respect to the $z$-axis, which is taken to be the symmetry axis of the MDC. Photon candidates are identified using showers in the electromagnetic calorimeter (EMC). The deposited energy of each shower must be more than 25~MeV in the barrel region $(|\!\cos\theta|<0.8)$ and more than 50~MeV in the end-cap region $(0.86<|\!\cos\theta|<0.92)$. To exclude showers that originate from charged tracks, the opening angle enclosed by the EMC shower and the position of the closest charged track at the EMC must be greater than 10 degrees as measured from the interaction point. To suppress electronic noise and showers unrelated to the event, the difference between the EMC time and the event start time is required to be within $[0, 700]$~ns. Particle identification (PID) for charged tracks combines measurements of the specific ionization energy loss in the MDC (d$E$/d$x$) and the flight time in the time-of-flight system (TOF) to form likelihoods $\mathcal{L}(h)$ $(h=p, K, \pi)$ for each hadron $h$ hypothesis. Tracks are identified as protons when the proton hypothesis has the greatest likelihood $(\mathcal{L}(p)>\mathcal{L}(\pi)$ and $\mathcal{L}(p)>\mathcal{L}(K))$, while tracks are identified as pions when the pion hypothesis has the greatest likelihood $(\mathcal{L}(\pi)>\mathcal{L}(K)$ and $\mathcal{L}(\pi)>\mathcal{L}(p))$.

The final state of  the reaction $J/\psi\too\Sigma^+\bar{\Sigma}^-$ with $\Sigma^{+}n\too\Lambda p$ is $pp\bar{p}\pi^{-}\gamma\gamma$. Candidate events must have four charged tracks with zero net charge and at least two photon candidates. We require that there are two protons, one anti-proton and one $\pi^-$. For the decay $\bar{\Sigma}^{-}\too\bar{p}\pi^0$ with $\pi^{0}\too\gamma\gamma$, the invariant mass of the two photons is required to be in the $\pi^0$ signal region $[0.115, 0.150]$~GeV/$c^2$, and the momenta of the photons are adjusted according to the results of  a one-constraint kinematic fit that constrains their invariant mass to the known $\pi^0$ mass. In this Letter, all known masses are taken from the PDG~\cite{pdg}. If there is more than one $\pi^0$ candidate in an event, only the one with the minimum value of $|M(\bar{p}\pi^0)-m_{\bar{\Sigma}^-}|$ is retained, where $m_{\bar{\Sigma}^-}$ is the known mass of the $\bar{\Sigma}^-$. The $\bar{\Sigma}^-$ signal region is defined as $-0.015$~GeV/$c^{2}<(M(\bar{p}\pi^0)-m_{\bar{\Sigma}^-})<0.010$~GeV/$c^{2}$. For the reaction $\Sigma^{+}n\too\Lambda p$ with subsequent decay $\Lambda\too p\pi^-$, we first perform a vertex fit to $p\pi^-$ combinations, of which the one with invariant mass closest to $m_{\Lambda}$ is taken as $\Lambda$ candidate, where $m_{\Lambda}$ is the known $\Lambda$ mass. The $\Lambda$ signal region is defined as $|M(p\pi^-)-m_{\Lambda}|<0.003$~GeV/$c^{2}$. Finally, a vertex fit is performed to the combination of the $\Lambda$ and the remaining $p$.

To select the  $J/\psi\too\Sigma^+\bar{\Sigma}^-$ signal events, the invariant mass of the system recoiling against the $\bar{\Sigma}^-$, $M_{\text{recoil}}(\bar{\Sigma}^-)$, is required to be in the $\Sigma^+$ signal region, defined as $[1.170, 1.205]$~GeV/$c^{2}$, where $M_{\text{recoil}}(\bar{\Sigma}^-)\equiv\sqrt{E^2_{\text{beam}}-|\vec{p}_{\bar{\Sigma}^-}c|^2}/c^2$, $E_{\text{beam}}$ is the beam energy, and $\vec{p}_{\bar{\Sigma}^-}$ is the measured momentum of the $\bar{\Sigma}^-$ candidate in the $\EE$ rest frame. The main background arises from the process $J/\psi\too\Sigma^+\bar{\Sigma}^-$, $\Sigma^{+}\too p\pi^0$, $\bar{\Sigma}^{-}\too\bar{p}\pi^0$. To suppress this contamination, the recoil masses of $\bar{\Sigma}^-p$ and $\bar{\Sigma}^-p_{\Lambda}$, $M_{\text{recoil}}(\bar{\Sigma}^-p)$ and $M_{\text{recoil}}(\bar{\Sigma}^-p_{\Lambda})$ are determined, where $p_{\Lambda}$ is the $p$ from $\Lambda$ candidate. If the missing energy is less than the missing momentum, we force the recoil-mass squared to be positive,  and set the recoil mass itself to be negative. The values of $M_{\text{recoil}}(\bar{\Sigma}^-p)$ or $M_{\text{recoil}}(\bar{\Sigma}^-p_{\Lambda})$ are expected to be around the known $\pi^0$ mass for this background, and so we require $M_{\text{recoil}}(\bar{\Sigma}^-p)<0$~GeV/$c^2$ and $M_{\text{recoil}}(\bar{\Sigma}^-p_{\Lambda})<0$~GeV/$c^2$ to suppress this background.

\begin{figure}[htbp]
\begin{center}
\begin{overpic}[width=0.36\textwidth]{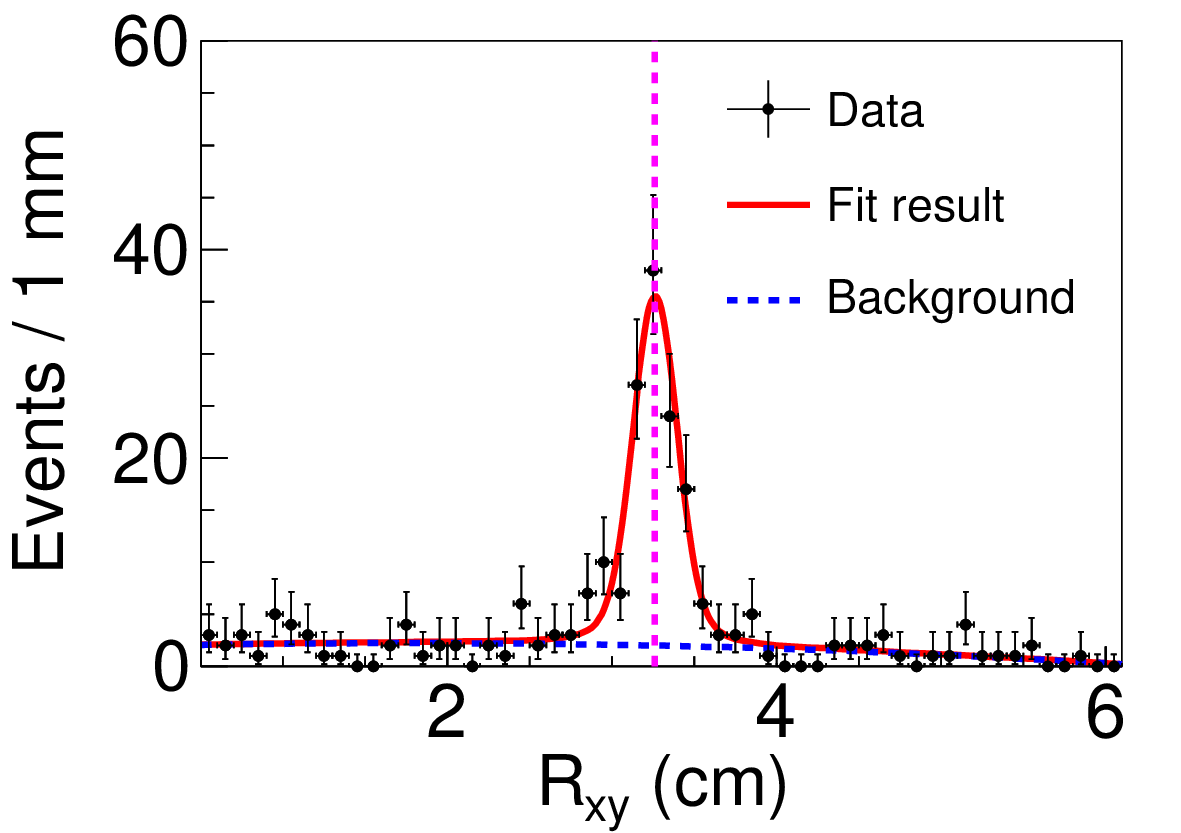}
\end{overpic}
\caption{Distribution of $R_{xy}$ in data with fit result superimposed, and the pink dashed line marks the center position of beam pipe. A clear excess is observed around the beam-pipe region, corresponding to the reaction $\Sigma^{+}n\too\Lambda p$ or $\Sigma^{+}n\too\Sigma^{0} p\too\gamma\Lambda p$.}
\label{fig:fit_result}
\end{center}
\end{figure}

Figure~\ref{fig:fit_result} shows the $R_{xy}$ distribution of the accepted candidates from data after final event selection, where $R_{xy}$ is the distance from the reconstructed $\Lambda p$ vertex to the $z$-axis. A clear excess is observed around the beam-pipe region, corresponding to the reaction $\Sigma^{+}n\too\Lambda p$ or $\Sigma^{+}n\too\Sigma^{0} p\too\gamma\Lambda p$ via interactions of $\Sigma^+$ with material in the beam pipe. A detailed study of the $J/\psi$ inclusive MC sample indicates that there is negligible peaking background contribution in this region.  Furthermore, no significant peak is found in the sideband region of the $M_{\text{recoil}}(\bar{\Sigma}^-)$ and $M(p\pi^-)$ distributions in data, where the $M_{\text{recoil}}(\bar{\Sigma}^-)$ sideband region is defined as $[1.100, 1.135]$ and $[1.240, 1.275]$~GeV/$c^{2}$, and the $M(p\pi^-)$ sideband region is defined as $0.009<|M(p\pi^-)-m_{\Lambda}|<0.015$~GeV/$c^{2}$. To determine the signal yield, an unbinned maximum likelihood fit is performed to the $R_{xy}$ distribution. We use the MC-simulated shape convolved by a Gaussian function with free parameters to describe the signal, and the background is described by a second-order Chebyshev polynomial function. The fit result is shown in Fig.~\ref{fig:fit_result}, and the fitted signal yield is $N^{\rm{total}}=126.2\pm13.4$. This signal yield is from the two processes $\Sigma^{+}n\too\Lambda p$ and $\Sigma^{+}n\too\Sigma^{0} p\too\gamma\Lambda p$, which cannot be distinguished in the $R_{xy}$ distribution. The selection efficiencies for the reactions $\Sigma^{+}n\too\Lambda p$ and $\Sigma^{+}n\too\Sigma^{0} p$ are $\epsilon_{\Lambda}=14.44\%$ and $\epsilon'_{\Sigma^0}=13.71\%$, respectively.

\begin{figure}[htbp]
\begin{center}
\begin{overpic}[width=0.36\textwidth]{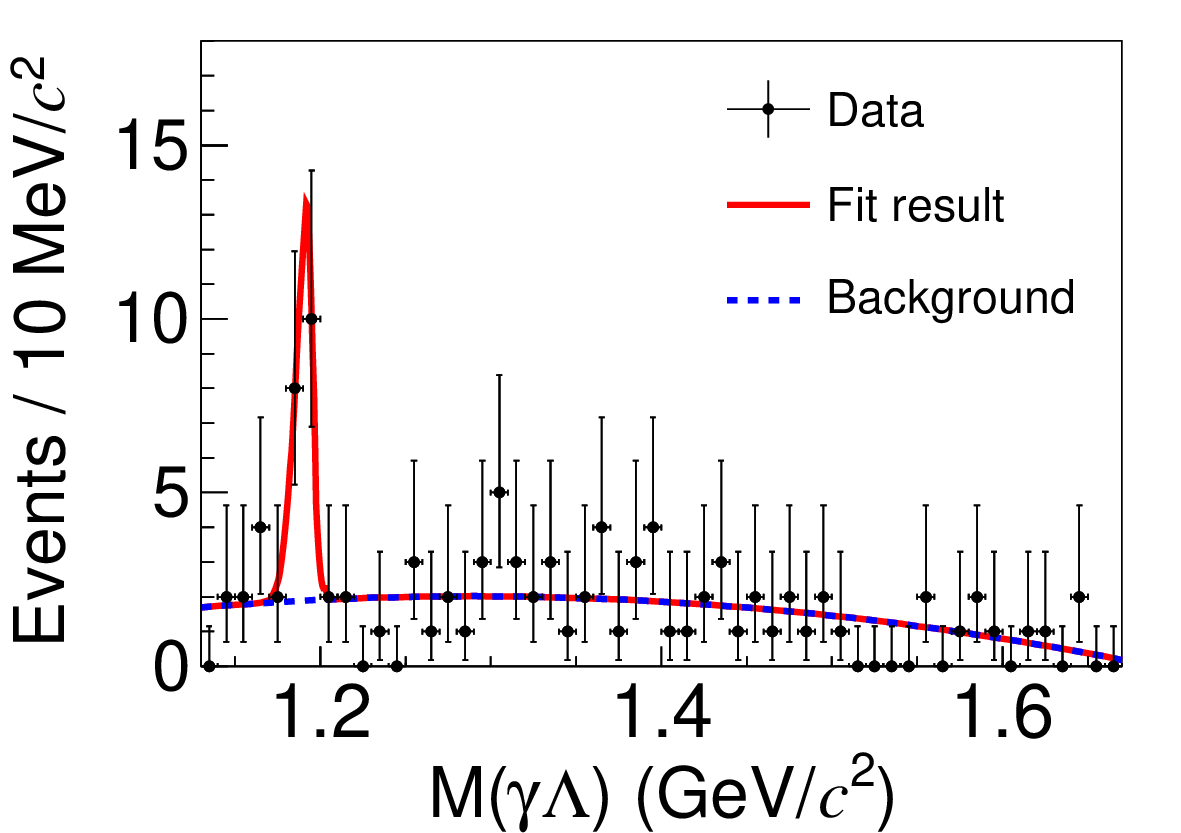}
\end{overpic}
\caption{Distribution of $M(\gamma\Lambda)$ in data with fit result superimposed, where the data refer to events which are in the signal region of the $R_{xy}$ distribution. Clear $\Sigma^0$ signals are seen, corresponding to the reaction $\Sigma^{+}n\too\Sigma^{0}p$.}
\label{fig:fit_result2}
\end{center}
\end{figure}

To separate the contributions of $\Sigma^{+}n\too\Lambda p$ and $\Sigma^{+}n\too\Sigma^{0} p$, we reconstruct the final state $\Sigma^0$ by requiring the presence of at least one additional photon. If there is more than one additional photon in an event, only the photon with maximum energy is taken as the photon from  the $\Sigma^0$ decay.  Figure~\ref{fig:fit_result2} shows the $M(\gamma\Lambda)$ distribution for the events in the beam-pipe signal region from data, where this region is defined as $[2.9, 3.6]$~cm. Clear $\Sigma^0$ signals are seen, corresponding to the reaction $\Sigma^{+}n\too\Sigma^{0}p$, and no significant peak is found in the $J/\psi$ inclusive MC sample and $R_{xy}$ sideband events from data, where the $R_{xy}$ sideband region is defined as $[1.6, 2.3]$ and $[4.2, 4.9]$~cm. To determine the signal yield, the same fit method as described above is applied to the $M(\gamma\Lambda)$ distribution. The fit, as presented in Fig.~\ref{fig:fit_result2}, yields $N_{\Sigma^0}=14.1\pm4.6$ signal events. The selection efficiency for the reaction $\Sigma^{+}n\too\Sigma^{0} p$ is $\epsilon_{\Sigma^0}=3.98\%$. Therefore, the number of signal events for the reaction $\Sigma^{+}n\too\Sigma^{0} p$ in the $R_{xy}$ distribution of Fig.~\ref{fig:fit_result} is determined to be $N_{\Sigma^0}\frac{\epsilon'_{\Sigma^0}}{\epsilon_{\Sigma^0}}=48.6\pm15.9$, and  the number of signal events for the reaction $\Sigma^{+}n\too\Lambda p$ in Fig.~\ref{fig:fit_result} is determined to be $N_{\Lambda}=N^{\rm{total}}-N_{\Sigma^0}\frac{\epsilon'_{\Sigma^0}}{\epsilon_{\Sigma^0}}=77.6\pm20.8$.

Using the same method as in Ref.~\cite{xiscatteringbes}, the cross-sections of the reactions $\Sigma^{+}+{^9\rm{Be}}\too\Lambda+p+X$ and $\Sigma^{+}+{^9\rm{Be}}\too\Sigma^{0}+p+X$ are determined as
\begin{linenomath*}
\begin{equation}
    \sigma(\Sigma^{+}+{^9\rm{Be}}\too\Lambda/\Sigma^{0}+p+X) = \frac{N_{\Lambda/\Sigma^{0}}} {\epsilon_{\Lambda/\Sigma^{0}} \cdot \mathcal{B} \cdot \mathcal{L}_{\rm{eff}} },
\end{equation}
\end{linenomath*}
where $\mathcal{B}$ is the product of the branching fractions of all intermediate resonances, defined as $\mathcal{B}\equiv\mathcal{B}(\bar{\Sigma}^{-}\too\bar{p}\pi^0)\mathcal{B}(\pi^0\too\gamma\gamma)\mathcal{B}(\Lambda\too p\pi^-)$, and $\mathcal{L}_{\rm{eff}}$ is the effective integrated luminosity of the $\Sigma^{+}$ flux produced from $J/\psi\too\Sigma^{+}\bar{\Sigma}^-$ and the distribution of target materials, as given by
\begin{linenomath*}
\begin{equation}
\small
    \mathcal{L}_{\rm{eff}} = \frac{\it{N}_{\it{J}/\psi}\mathcal{B}_{\it{J}/\psi}}{\rm{2}+\frac{\rm{2}}{\rm{3}}\alpha} \int_{a}^{b}\int_{\rm{0}}^{\pi} [(\rm{1}+\alpha \rm{cos}^2\theta) \rm{exp}(-\frac{\it{x}}{\rm{sin}\theta \it{\beta\gamma L}})\it{N(x)C(x)}] \rm{d}\theta \rm{d}\it{x}.
\end{equation}
\end{linenomath*}
In this formula, the angular distribution of the $\Sigma^{+}$ flux, the attenuation of the $\Sigma^{+}$ flux, the number of target nuclei, and the weight of different target materials are taken into account. $N_{J/\psi}$ is the total number of $J/\psi$ events~\cite{totalnumber}, $\mathcal{B}_{J/\psi}$ is the branching fraction of $J/\psi\too\Sigma^{+}\bar{\Sigma}^{-}$, $\alpha$ is the parameter that governs  the angular distribution of $J/\psi\too\Sigma^{+}\bar{\Sigma}^{-}$ decays~\cite{alpha}, $\beta\gamma\equiv \frac{\sqrt{E_{\rm{beam}}^{\rm{2}}-m_{\rm{\Sigma^{+}}}^{\rm{2}}c^{\rm{4}}}}{m_{\rm{\Sigma^{+}}}c^{\rm{2}}}$ is the ratio of the momentum and the mass of the $\Sigma^{+}$, $L\equiv c\tau$, where $\tau$ is the mean lifetime of the $\Sigma^{+}$, $N(x)$ is the number of target nuclei per unit volume, $a$ and $b$ are the distances from the inner and outer surfaces of the beam pipe to the $z$-axis, and $\theta$ and $x$ are the angle and distance to the $z$-axis~\cite{xiscatteringbes}. The beam pipe can be regarded as infinitely long with respect to the product $\beta\gamma L$ of $\Sigma^{+}$. $C(x)$ is the cross-section ratio relative to $\sigma(\Sigma^{+}+{^9\rm{Be}}\too\Lambda/\Sigma^{0}+p+X)$, where the reaction is assumed to be dominated by the interaction of a $\Sigma^{+}$ baryon with a single neutron on the nucleus surface~\cite{ratio1, ratio2, ratio3, ratio4, ratio5}, which can be found in Ref.~\cite{xiscatteringbes}. The measured cross-sections of the two reactions are listed in Table~\ref{tab:crosssection}.

\begin{table}[htbp]
\begin{center}
\caption{Measured cross-sections for the reactions $\Sigma^{+}+{^9\rm{Be}}\rightarrow\Lambda+p+X$ and $\Sigma^{+}+{^9\rm{Be}}\rightarrow\Sigma^{0}+p+X$ at the $\Sigma^+$ momentum of about 0.992 GeV/$c$.}
\label{tab:crosssection}
\begin{tabular}{cc}
  \hline
  \hline
  Parameter & Cross-section (mb) \\
  \hline
  \ \ \ $\sigma(\Sigma^{+}+{^9\rm{Be}}\rightarrow\Lambda+p+X)$ \ \ \ & \ \ \ $45.2\pm12.1_{\rm{stat}}\pm7.2_{\rm{sys}}$ \ \ \ \\
  $\sigma(\Sigma^{+}+{^9\rm{Be}}\rightarrow\Sigma^{0}+p+X)$ & $29.8\pm9.7_{\rm{stat}}\pm6.9_{\rm{sys}}$  \\
  \hline
  \hline
\end{tabular}
\end{center}
\end{table}

The sources of systematic uncertainties on the measured cross-sections are now discussed.

The uncertainty in the reconstruction or PID efficiency of charged tracks, and the reconstruction efficiency of photons is $1\%$ for each track or photon~\cite{track}. The uncertainty associated with the track-number requirement is studied with a   control sample $J/\psi\too\Lambda\bar{\Lambda}\too p\pi^-\bar{p}\pi^+$ events.

In MC simulation, we take the momentum of the neutron in the nucleus to be zero, as the Fermi-momentum is very small compared to the momentum of the incident $\Sigma^+$. However, this approximation leads to a discrepancy in the distribution of $(\Lambda/\Sigma^{0}+p)$ momentum $P(\Lambda/\Sigma^{0}+p)$ for data and MC simulation. The change of $P(\Lambda/\Sigma^{0}+p)$ due to the Fermi-momentum for most events in data is within $\pm0.1$~GeV/$c$. Therefore, to estimate the uncertainty from $P(\Lambda/\Sigma^{0}+p)$, we vary the momentum of the free neutron by $\pm0.1$~GeV/$c$ along the direction of the incident $\Sigma^+$ in the generated signal MC, and take the larger difference as the uncertainty. The distribution of $M(\Lambda p/\Sigma^{0}p)$ is assumed to be flat in the baseline signal MC. To obtain the uncertainty from the $M(\Lambda p/\Sigma^{0}p)$ distribution, the difference in the efficiency between the baseline signal MC and the MC weighted according to the distribution of signal events in data is assigned as the systematic uncertainty. The reactions $\Sigma^{+}n\too\Lambda p$ and $\Sigma^{+}n\too\Sigma^{0}p$ are simulated with a uniform angular distribution over the phase space to estimate the baseline efficiency. The weighted efficiency of signal events is calculated based on real data, and the difference between the baseline and weighted efficiencies is taken as the uncertainty.

The uncertainty associated with the fit procedure arises from the assumed background shape and the fit range. The uncertainty from the background shape is estimated by changing a second-order polynomial function to a third-order one, and the uncertainty from the fit range is obtained by varying the limit of the fit range by $\pm0.1$~cm or $\pm0.01$~GeV/$c^{2}$. The uncertainty from the total number of $J/\psi$ events is reported in Ref.~\cite{totalnumber}. The uncertainty of the branching fractions $\mathcal{B}$ and $\mathcal{B}_{J/\psi}$ is taken from the PDG~\cite{pdg}. To estimate the uncertainties from the angular distribution of $J/\psi\too\Sigma^+\bar{\Sigma}^-$ and the $\Sigma^+$ mean lifetime, we vary the angular-distribution parameter $\alpha$ and the mean lifetime by $\pm1\sigma$. The uncertainty from the position of the $\EE$ interaction point is obtained by changing the integral range by $\pm0.1$~cm, which is from $(a, b)$ to $(a+0.1, b+0.1)$ or $(a-0.1, b-0.1)$, and the larger difference in the result is taken as the uncertainty.

When determining the cross-sections, we assume that the interactions involve individual neutrons on the surface of the nucleus. To estimate the uncertainty from this assumption, we adopt an alternative extreme assumption that the cross-sections are proportional to the number of neutrons in the nucleus~\cite{ratio1, ratio2, ratio3, ratio4, ratio5}, and the difference in the results between the two  assumptions is taken as the uncertainty. The systematic uncertainties from the mass windows, $M_{\text{recoil}}(\bar{\Sigma}^{-}p)/M_{\text{recoil}}(\bar{\Sigma}^{-}p_{\Lambda})$ requirements and $R_{xy}$ requirement are assessed using the Barlow test method~\cite{lambdascatteringbes}, and are found to be negligible.

A summary of the systematic uncertainties is presented in Table~\ref{tab:sumerror}. The total systematic uncertainty is obtained by adding all the individual components in quadrature.
\begin{table}[htbp]
\begin{center}
\caption{Summary of relative systematic uncertainties for measured cross-sections (in $\%$). $\sigma_{\Lambda}$ represents the cross-section $\sigma(\Sigma^{+}+{^9\rm{Be}}\rightarrow\Lambda+p+X)$, and $\sigma_{\Sigma^0}$ represents the cross-section $\sigma(\Sigma^{+}+{^9\rm{Be}}\rightarrow\Sigma^{0}+p+X)$.}
\label{tab:sumerror}
\begin{tabular}{ccc}
  \hline
  \hline
  Source & $\sigma_{\Lambda}$ & $\sigma_{\Sigma^0}$ \\
  \hline
  Tracking                                                       & 4.0  & 4.0  \\
  Photon                                                         & 2.0  & 3.0  \\
  PID                                                            & 4.0  & 4.0  \\
  Track number                                                   & 2.2  & 2.2  \\
  $(\Lambda/\Sigma^{0}+p)$ momentum                              & 6.0  & 12.5 \\
  $M(\Lambda p/\Sigma^{0}p)$ distribution                        & 2.4  & 2.8  \\
  \ \ Angular distribution of $\Sigma^{+}n\too\Lambda p/\Sigma^{0}p$ \ \ & \ \ 6.5 \ \ & \ \ 14.3 \ \ \\
  Fit procedure                                                  & 5.4  & 4.4  \\
  Number of $J/\psi$                                             & 0.4  & 0.4  \\
  Branching fractions                                            & 3.5  & 3.5  \\
  Angular distribution of $J/\psi\too\Sigma^+\bar{\Sigma}^-$     & 0.1  & 0.1  \\
  $\Sigma^+$ mean lifetime                                       & 0.7  & 0.7  \\
  $e^+e^-$ interaction point                                     & 6.0  & 6.0  \\
  Cross-section ratios                                           & 7.3  & 7.3  \\
  \hline
  Sum                                                            & 16.0 & 23.2 \\
  \hline
  \hline
\end{tabular}
\end{center}
\end{table}

In summary, using $(1.0087\pm0.0044)\times10^{10}$ $J/\psi$ events collected with the BESIII detector operating at the BEPCII storage ring, the reactions $\Sigma^{+}n\too\Lambda p$ and $\Sigma^{+}n\too\Sigma^{0}p$ have been studied for the first time, where $\Sigma^{+}$ is generated from the process $J/\psi\rightarrow\Sigma^{+}\bar{\Sigma}^-$ and $n$ originates from materials in the beam pipe. The cross-sections of these two reactions are measured to be $\sigma(\Sigma^{+}+{^9\rm{Be}}\too\Lambda+p+X)=(45.2\pm12.1_{\rm{stat}}\pm7.2_{\rm{sys}})$~mb and $\sigma(\Sigma^{+}+{^9\rm{Be}}\too\Sigma^{0}+p+X)=(29.8\pm9.7_{\rm{stat}}\pm6.9_{\rm{sys}})$~mb at a $\Sigma^{+}$ momentum of $0.992$~GeV/$c$, within a range of $\pm0.015$~GeV/$c$. Assuming the effective number of reaction neutrons in a beryllium nucleus to be approximately $3$~\cite{introduction8}, the cross-sections of $\Sigma^{+}n\too\Lambda p$ and $\Sigma^{+}n\too\Sigma^{0}p$ for a single neutron are determined to be $\sigma(\Sigma^{+}n\too\Lambda p)=(15.1\pm4.0_{\rm{stat}}\pm2.4_{\rm{sys}})$~mb and $\sigma(\Sigma^{+}n\too\Sigma^{0}p)=(9.9\pm3.2_{\rm{stat}}\pm2.3_{\rm{sys}})$~mb, respectively, which is in agreement with theoretical predictions based on the leading order covariant chiral effective field theory in Ref.~\cite{theory6}. It is worth mentioning that anyone can use an effective number of reaction neutrons they deem more accurate to derive $\sigma(\Sigma^{+}n\too\Lambda p/\Sigma^0 p)$ from the measured $\sigma(\Sigma^{+}+{^9\rm{Be}}\too\Lambda/\Sigma^0+p+X)$. Furthermore, these cross-sectional measurements can be used to verify the reliability of other theoretical models.

Despite the lifetime of $\Sigma^+$ is three times shorter than $\Lambda/\bar{\Lambda}$ and $\Xi^0$ those were previously studied at BESIII~\cite{xiscatteringbes, lambdabarscatteringbes, lambdascatteringbes}, it is still feasible to study the $\Sigma^+$-nucleon reaction using the beam pipe as the target material. Consequently, this study marks the first exploration of $\Sigma^+$-nucleon scattering at an electron-positron collider, and validates the feasibility of using this novel method to study the shorter lifetime $\Sigma^+$. Especially, in all these studies of hyperon-nucleon reactions at BESIII, only the reaction $\Sigma^{+}n\too\Lambda p$ is exoenergetic, while the reaction $\Sigma^{+}n\too\Sigma^{0}p$ and other studied inelastic scattering processes $\Xi^0n\too\Xi^-p$ and $\Lambda p\too\Sigma^+n$~\cite{xiscatteringbes, lambdascatteringbes} are endergonic. Different reactions will affect the momentum of hyperons in neutron stars, and further influence the equation of state, this constitutes an important factor in resolving the ``hyperon puzzle" of neutron stars. Furthermore, these measured results offer vital insights for studying the $\Lambda$N-$\Sigma$N coupling, and serve as a significant element in enhancing the comprehensive understanding of $\Sigma$-nucleon interaction. With more statistics in future super tau-charm facilities~\cite{super1, super2}, the momentum-dependent $\Sigma$-nucleon cross-section distribution can also be studied using the $\Sigma$ sources from multibody decays of $J/\psi$ or other charmonia, such as $J/\psi\too\bar{\Lambda}\pi^-\Sigma^+$+c.c. and $\psi(2S)\rightarrow\Sigma^{+}\bar{\Sigma}^-$.

The BESIII Collaboration thanks the staff of BEPCII and the IHEP computing center for their strong support. This work is supported in part by National Key R\&D Program of China under Contracts Nos. 2020YFA0406300, 2020YFA0406400, 2023YFA1606000; National Natural Science Foundation of China (NSFC) under Contracts Nos. 12375071, 11635010, 11735014, 11935015, 11935016, 11935018, 12025502, 12035009, 12035013, 12061131003, 12192260, 12192261, 12192262, 12192263, 12192264, 12192265, 12221005, 12225509, 12235017, 12361141819; Natural Science Foundation of Henan under Contract No. 242300421163; the Chinese Academy of Sciences (CAS) Large-Scale Scientific Facility Program; the CAS Center for Excellence in Particle Physics (CCEPP); Joint Large-Scale Scientific Facility Funds of the NSFC and CAS under Contract No. U1832207; CAS under Contract No. YSBR-101; 100 Talents Program of CAS; The Institute of Nuclear and Particle Physics (INPAC) and Shanghai Key Laboratory for Particle Physics and Cosmology; German Research Foundation DFG under Contract No. FOR5327; Istituto Nazionale di Fisica Nucleare, Italy; Knut and Alice Wallenberg Foundation under Contracts Nos. 2021.0174, 2021.0299; Ministry of Development of Turkey under Contract No. DPT2006K-120470; National Research Foundation of Korea under Contract No. NRF-2022R1A2C1092335; National Science and Technology fund of Mongolia; National Science Research and Innovation Fund (NSRF) via the Program Management Unit for Human Resources \& Institutional Development, Research and Innovation of Thailand under Contract No. B50G670107; Polish National Science Centre under Contract No. 2019/35/O/ST2/02907; Swedish Research Council under Contract No. 2019.04595; The Swedish Foundation for International Cooperation in Research and Higher Education under Contract No. CH2018-7756; U. S. Department of Energy under Contract No. DE-FG02-05ER41374.


\begin{thebibliography}{**}

\bibitem{introduction1} B.~Sechi-Zorn, B.~Kehoe, J.~Twitty and R.~A.~Burnstein, \href{https://doi.org/10.1103/PhysRev.175.1735}{\color{blue} Phys. Rev. \textbf{175}, 1735 (1968)}.

\bibitem{introduction2} G.~Alexander, U.~Karshon, A.~Shapira, G.~Yekutieli, R.~Engelmann, H.~Filthuth and W.~Lughofer, \href{https://doi.org/10.1103/PhysRev.173.1452}{\color{blue} Phys. Rev. \textbf{173}, 1452 (1968)}.

\bibitem{introduction3} R.~A.~Muller, \href{https://doi.org/10.1016/0370-2693(72)90757-5}{\color{blue} Phys. Lett. B \textbf{38}, 123 (1972)}.

\bibitem{introduction4} J.~M.~Hauptman, J.~A.~Kadyk and G.~H.~Trilling, \href{https://doi.org/10.1016/0550-3213(77)90222-X}{\color{blue} Nucl. Phys. B \textbf{125}, 29 (1977)}.

\bibitem{introduction5} M.~Bourquin and J.~P.~Repellin, \href{https://doi.org/10.1016/0370-1573(84)90041-3}{\color{blue} Phys. Rept. \textbf{114}, 99 (1984)}.

\bibitem{introduction6} S.~F.~Biagi \textit{et al.}, \href{https://doi.org/10.1007/BF01566759}{\color{blue} Z. Phys. C \textbf{34}, 187 (1987)}.

\bibitem{introduction7} J.~K.~Ahn \textit{et al.} [E224 Collaboration], \href{https://doi.org/10.1016/S0375-9474(97)81462-5}{\color{blue} Nucl. Phys. A \textbf{625}, 231 (1997)}.

\bibitem{introduction8} J.~K.~Ahn \textit{et al.}, \href{https://doi.org/10.1016/j.physletb.2005.12.057}{\color{blue} Phys. Lett. B \textbf{633}, 214 (2006)}.

\bibitem{introduction9} J.~Rowley \textit{et al.} [CLAS Collaboration], \href{https://doi.org/10.1103/PhysRevLett.127.272303}{\color{blue} Phys. Rev. Lett. \textbf{127}, 272303 (2021)}.

\bibitem{neutronstar1} I.~Vida\~na, \href{https://doi.org/10.1016/j.nuclphysa.2013.01.015}{\color{blue} Nucl. Phys. A \textbf{914}, 367 (2013)}.

\bibitem{neutronstar2} D.~Chatterjee and I.~Vida\~na, \href{https://doi.org/10.1140/epja/i2016-16029-x}{\color{blue} Eur. Phys. J. A \textbf{52}, 29 (2016)}.

\bibitem{neutronstar3} I.~Vida\~na, \href{https://doi.org/10.1098/rspa.2018.0145}{\color{blue} Proc. Roy. Soc. Lond. A \textbf{474}, 0145 (2018)}.

\bibitem{neutronstar4} L.~Tolos and L.~Fabbietti, \href{https://doi.org/10.1016/j.ppnp.2020.103770}{\color{blue} Prog. Part. Nucl. Phys. \textbf{112}, 103770 (2020)}.

\bibitem{lambda-sigma1} E.~Hiyama and K.~Nakazawa, \href{https://doi.org/10.1146/annurev-nucl-101917-021108}{\color{blue} Ann. Rev. Nucl. Part. Sci. \textbf{68}, 131 (2018)}.

\bibitem{lambda-sigma2} I.~Vidana, D.~Logoteta, C.~Providencia, A.~Polls and I.~Bombaci, \href{https://doi.org/10.1209/0295-5075/94/11002}{\color{blue} EPL \textbf{94}, 11002 (2011)}.

\bibitem{lambda-sigma3} B.~F.~Gibson, A.~Goldberg and M.~S.~Weiss, \href{https://doi.org/10.1103/PhysRevC.6.741}{\color{blue} Phys. Rev. C \textbf{6}, 741 (1972)}.

\bibitem{lambda-sigma4} J.~Haidenbauer, U.~G.~Mei\ss{}ner and A.~Nogga, \href{https://doi.org/10.1140/epja/s10050-020-00100-4}{\color{blue} Eur. Phys. J. A \textbf{56}, 91 (2020)}.

\bibitem{lambda-sigma5} S.~Acharya \textit{et al.} [ALICE Collaboration], \href{https://doi.org/10.1016/j.physletb.2022.137272}{\color{blue} Phys. Lett. B \textbf{833}, 137272 (2022)}.

\bibitem{lambda-sigma6} F.~Hildenbrand, S.~Elhatisari, Z.~Ren and U.~G.~Mei\ss{}ner, \href{https://doi.org/10.1140/epja/s10050-024-01427-y}{\color{blue} Eur. Phys. J. A \textbf{60}, 215  (2024)}.

\bibitem{csb1} T.~O.~Yamamoto \textit{et al.} [J-PARC E13 Collaboration], \href{https://doi.org/10.1103/PhysRevLett.115.222501}{\color{blue} Phys. Rev. Lett. \textbf{115}, 222501 (2015)}.

\bibitem{csb2} D.~Gazda and A.~Gal, \href{https://doi.org/10.1103/PhysRevLett.116.122501}{\color{blue} Phys. Rev. Lett. \textbf{116}, 122501 (2016)}.

\bibitem{csb3} J.~Haidenbauer, U.~G.~Mei\ss{}ner and A.~Nogga, \href{https://doi.org/10.1007/s00601-021-01684-3}{\color{blue} Few Body Syst. \textbf{62}, 105 (2021)}.

\bibitem{theory1} Z.~Y.~Zhang, Y.~W.~Yu, P.~N.~Shen, L.~R.~Dai, A.~Faessler and U.~Straub, \href{https://doi.org/10.1016/S0375-9474(97)00033-X}{\color{blue} Nucl. Phys. A \textbf{625}, 59 (1997)}.

\bibitem{theory2} Y.~Fujiwara, C.~Nakamoto and Y.~Suzuki, \href{https://doi.org/10.1103/PhysRevLett.76.2242}{\color{blue} Phys. Rev. Lett. \textbf{76}, 2242 (1996)}.

\bibitem{theory3} Y.~Fujiwara, C.~Nakamoto and Y.~Suzuki, \href{https://doi.org/10.1103/PhysRevC.54.2180}{\color{blue} Phys. Rev. C \textbf{54}, 2180 (1996)}.

\bibitem{theory4} T.~A.~Rijken, V.~G.~J.~Stoks and Y.~Yamamoto, \href{https://doi.org/10.1103/PhysRevC.59.21}{\color{blue} Phys. Rev. C \textbf{59}, 21 (1999)}.

\bibitem{theory5} J.~Haidenbauer and U.~G.~Mei\ss{}ner, \href{https://doi.org/10.1103/PhysRevC.72.044005}{\color{blue} Phys. Rev. C \textbf{72}, 044005 (2005)}.

\bibitem{theory6} J.~Song, Z.~W.~Liu, K.~W.~Li and L.~S.~Geng, \href{https://doi.org/10.1103/PhysRevC.105.035203}{\color{blue} Phys. Rev. C \textbf{105}, 035203 (2022)}.

\bibitem{theory7} J.~Haidenbauer and U.~G.~Mei\ss{}ner, \href{https://doi.org/10.1016/j.physletb.2022.137074}{\color{blue} Phys. Lett. B \textbf{829}, 137074 (2022)}.

\bibitem{theory8} J.~Haidenbauer, U.~G.~Mei\ss{}ner, A.~Nogga and H.~Le, \href{https://doi.org/10.1140/epja/s10050-023-00960-6}{\color{blue} Eur. Phys. J. A \textbf{59}, 63 (2023)}.

\bibitem{theory9} V.~G.~J.~Stoks and T.~A.~Rijken, \href{https://doi.org/10.1103/PhysRevC.59.3009}{\color{blue} Phys. Rev. C \textbf{59}, 3009 (1999)}.

\bibitem{theory10} Q.~Liu and I.~Low, \href{https://doi.org/10.1016/j.physletb.2024.138899}{\color{blue} Phys. Lett. B \textbf{856}, 138899 (2024)}.

\bibitem{sigma1} G.~R.~Charlton, J.~Badier, E.~Barrelet, I.~Makarovisch, J.~Pernegr, J.~R.~Hubbard, A.~Leveque, C.~Louedec, L.~Moscoso and D.~Revel, \href{https://doi.org/10.1016/0370-2693(70)90454-5}{\color{blue} Phys. Lett. B \textbf{32}, 720 (1970)}.

\bibitem{sigma2} F.~Eisele, H.~Filthuth, W.~Foehlisch, V.~Hepp and G.~Zech, \href{https://doi.org/10.1016/0370-2693(71)90053-0}{\color{blue} Phys. Lett. B \textbf{37}, 204 (1971)}.

\bibitem{sigma3} S.~F.~Biagi \textit{et al.}, \href{https://doi.org/10.1016/0550-3213(81)90089-4}{\color{blue} Nucl. Phys. B \textbf{186}, 1 (1981)}.

\bibitem{sigma4} M.~I.~Adamovich \textit{et al.} [WA89 Collaboration], \href{https://doi.org/10.1007/s100520050631}{\color{blue} Eur. Phys. J. C \textbf{11}, 271 (1999)}.

\bibitem{sigma5} R.~Vogt and T.~D.~Gutierrez, \href{https://doi.org/10.1016/j.nuclphysa.2003.07.003}{\color{blue} Nucl. Phys. A \textbf{726}, 134 (2003)}.

\bibitem{sigma6} K.~Miwa \textit{et al.} [J-PARC E40 Collaboration], \href{https://doi.org/10.1103/PhysRevC.104.045204}{\color{blue} Phys. Rev. C \textbf{104}, 045204 (2021)}.

\bibitem{sigma7} T.~Nanamura \textit{et al.} [J-PARC E40 Collaboration], \href{https://doi.org/10.1093/ptep/ptac101}{\color{blue} PTEP \textbf{2022}, 093D01 (2022)}.

\bibitem{sigma8} K.~Miwa \textit{et al.} [J-PARC E40 Collaboration], \href{https://doi.org/10.1103/PhysRevLett.128.072501}{\color{blue} Phys. Rev. Lett. \textbf{128}, 072501 (2022)}.

\bibitem{bepcii} C.~H.~Yu \textit{et al.}, Proceedings of IPAC2016, Busan, Korea, 2016, doi: \href{https://doi.org/10.18429/JACoW-IPAC2016-TUYA01}{\color{blue} 10.18429/JACoW-IPAC2016-TUYA01}.

\bibitem{besiii} M.~Ablikim \textit{et al.} [BESIII Collaboration], \href{https://linkinghub.elsevier.com/retrieve/pii/S0168900209023870}{\color{blue} Nucl. Instrum. Meth. A \textbf{614}, 345 (2010)}.

\bibitem{xiscatteringbes} M.~Ablikim \textit{et al.} [BESIII Collaboration], \href{https://doi.org/10.1103/PhysRevLett.130.251902}{\color{blue} Phys. Rev. Lett. \textbf{130}, 251902 (2023)}.

\bibitem{lambdabarscatteringbes} M.~Ablikim \textit{et al.} [BESIII Collaboration], \href{https://doi.org/10.1103/PhysRevLett.132.231902}{\color{blue} Phys. Rev. Lett. \textbf{132}, 231902 (2024)}.

\bibitem{lambdascatteringbes} M.~Ablikim \textit{et al.} [BESIII Collaboration], \href{https://doi.org/10.1103/PhysRevC.109.L052201}{\color{blue} Phys. Rev. C \textbf{109}, L052201 (2024)}.

\bibitem{totalnumber} M.~Ablikim \textit{et al.} [BESIII Collaboration], \href{https://doi.org/10.1088/1674-1137/ac5c2e}{\color{blue} Chin. Phys. C \textbf{46}, 074001 (2022)}.

\bibitem{ratio1} M.~Astrua, E.~Botta, T.~Bressani, D.~Calvo, C.~Casalegno, A.~Feliciello, A.~Filippi, S.~Marcello, M.~Agnello and F.~Iazzi, \href{https://doi.org/10.1016/S0375-9474(01)01252-0}{\color{blue} Nucl. Phys. A \textbf{697}, 209 (2002)}.

\bibitem{geant4} S.~Agostinelli \textit{et al.} [GEANT4 Collaboration], \href{https://doi.org/10.1016/S0168-9002(03)01368-8}{\color{blue} Nucl. Instrum. Meth. A \textbf{506}, 250 (2003)}.

\bibitem{display} K.~X.~Huang, Z.~J.~Li, Z.~Qian, J.~Zhu, H.~Y.~Li, Y.~M.~Zhang, S.~S.~Sun and Z.~Y.~You, \href{https://doi.org/10.1007/s41365-022-01133-8}{\color{blue} Nucl. Sci. Tech. \textbf{33}, 142 (2022)}.

\bibitem{KKMC} S.~Jadach, B.~F.~L.~Ward and Z.~Was, \href{https://doi.org/10.1103/PhysRevD.63.113009}{\color{blue} Phys. Rev. D \textbf{63}, 113009 (2001)}; \href{https://doi.org/10.1016/S0010-4655(00)00048-5}{\color{blue} Comput. Phys. Commun. \textbf{130}, 260 (2000)}.

\bibitem{ref:evtgen} D.~J.~Lange, \href{https://doi.org/10.1016/S0168-9002(01)00089-4}{\color{blue} Nucl. Instrum. Meth. A \textbf{462}, 152 (2001)}; R.~G.~Ping, \href{https://doi.org/10.1088/1674-1137/32/8/001}{\color{blue} Chin. Phys. C \textbf{32}, 599 (2008)}.

\bibitem{pdg} S.~Navas \textit{et al.} [Particle Data Group], \href{https://doi.org/10.1103/PhysRevD.110.030001}{\color{blue} Phys. Rev. D \textbf{110}, 030001 (2024)}.

\bibitem{ref:lundcharm} J.~C.~Chen, G.~S.~Huang, X.~R.~Qi, D.~H.~Zhang and Y.~S.~Zhu, \href{https://doi.org/10.1103/PhysRevD.62.034003}{\color{blue} Phys. Rev. D \textbf{62}, 034003 (2000)}; R.~L.~Yang, R.~G.~Ping and H.~Chen, \href{https://doi.org/10.1088/0256-307X/31/6/061301}{\color{blue} Chin. Phys. Lett. \textbf{31}, 061301 (2014)}.

\bibitem{photos} E.~Richter-Was, \href{https://doi.org/10.1016/0370-2693(93)90062-M}{\color{blue} Phys. Lett. B \textbf{303}, 163 (1993)}.

\bibitem{alpha} M.~Ablikim \textit{et al.} [BESIII Collaboration], \href{https://doi.org/10.1103/PhysRevLett.125.052004}{\color{blue} Phys. Rev. Lett. \textbf{125}, 052004 (2020)}.

\bibitem{ratio2} D.~S.~Barton \textit{et al.}, \href{https://doi.org/10.1103/PhysRevD.27.2580}{\color{blue} Phys. Rev. D \textbf{27}, 2580 (1983)}.

\bibitem{ratio3} M.~I.~Adamovich \textit{et al.} [WA89 Collaboration], \href{https://doi.org/10.1007/s002880050524}{\color{blue} Z. Phys. C \textbf{76}, 35 (1997)}.

\bibitem{ratio4} E.~Botta, \href{https://doi.org/10.1016/S0375-9474(01)01157-5}{\color{blue} Nucl. Phys. A \textbf{692}, 39 (2001)}.

\bibitem{ratio5} T.~G.~Lee and C.~Y.~Wong, \href{https://doi.org/10.1103/PhysRevC.97.054617}{\color{blue} Phys. Rev. C \textbf{97}, 054617 (2018)}.

\bibitem{track} M.~Ablikim \textit{et al.} [BESIII Collaboration], \href{https://doi.org/10.1103/PhysRevD.104.072007}{\color{blue} Phys. Rev. D \textbf{104}, 072007 (2021)}.

\bibitem{super1} A.~E.~Bondar \textit{et al.} [Charm-Tau Factory Collaboration], \href{https://doi.org/10.1134/S1063778813090032}{\color{blue} Phys. Atom. Nucl. \textbf{76}, 1072 (2013)}.

\bibitem{super2} M.~Achasov \textit{et al.}, \href{https://doi.org/10.1007/s11467-023-1333-z}{\color{blue} Front. Phys. \textbf{19}, 14701 (2024)}.

\end{thebibliography}
\end{document}